\documentclass[superscriptaddress,runinaddress,showpacs,showkeys,aps,prd,floatfix,onecolumn]{revtex4}
\usepackage{amssymb,amsmath,epsfig}
\usepackage{amsmath,graphicx}
\usepackage{palatino}
\usepackage{changes}

\usepackage[colorlinks=true,
            linkcolor=blue,
            urlcolor=blue,
            citecolor=blue]{hyperref}
\begin{document}

\title{Weak Deflection angle and Shadow by Tidal Charged Black Hole}

\author{Wajiha Javed}
\email{wajiha.javed@ue.edu.pk; wajihajaved84@yahoo.com} 
\affiliation{Division of Science and Technology, University of Education,Township Campus, Lahore-54590, Pakistan}

\author{Ali Hamza}
\email{alihamza.ahg@gmail.com}
\affiliation{Division of Science and Technology, University of Education,Township Campus, Lahore-54590, Pakistan}

\author{Ali {\"O}vg{\"u}n}
\email{ali.ovgun@emu.edu.tr}\homepage{https://aovgun.weebly.com} 

\affiliation{Physics Department, Eastern Mediterranean
University, Famagusta, 99628 North Cyprus via Mersin 10, Turkey.}

\begin{abstract}
In this article, we calculate the deflection angle of
tidal charged black hole (TCBH) in weak field limits. First we obtain the Gaussian optical curvature and then apply the Gauss-Bonnet theorem on it. With the help of Gibbons-Werner method, we are able to calculate the light's deflection angle by TCBH in weak field limits. After calculating the deflection angle of light, we check the graphical behavior of TCBH. Moreover, we further find the light's deflection angle in the presence of plasma medium and also check the graphical behavior in the presence of plasma medium. Moreover, we investigate the shadow of TCBH.For calculating the shadow, we first find the null geodesics around the TCBH and then find its shadow radius. We also obtain TCBH's shadow in the plasma medium. Hence, we discuss the shadow of the TCBH using the $M87^{*}$ parameters announced by the Event Horizon Telescope.
\end{abstract}


\date{\today}
 \maketitle

\section{Introduction}
Einstein's theory of general relativity (GR) is a wonderful gravity theory which was developed in 1916. Einstein intelligently anticipated  the presence of gravitational waves and gravitational lensing in his theory of GR \cite{J1 1}. Black Holes (BHs) are very interesting and captivating objects in the universe. It is assumed that various types of BHs live in the universe and experimentally confirmed by the Laser Interferometer Gravitational-Wave Observatory (LIGO) experiment in 2015 \cite{6 1}, after that EHT Collaboration has shown the existence of the black hole by its shadow in 2019 \cite{Akiyama:2019cqa}. The BH's physics is very important because it plays a vital role in discovery of gravitational wave \cite{6 1} and further BH's physics is used for understanding entropy and the information paradox \cite{6 2}. It is also used in an interesting aspects of gravitational lensing. As indicated by Einstein, light bends around a massive object, such as a black hole, causing it to act as a lens for the things that lie behind it. Strong lensing generates curves and rings, for example, the Einstein's ring, while weak lensing gravity does not create the image of a distant galaxy, but it still produces a measurable useful effect. Moreover, weak lensing provides an independent measurement of dark energy, the substance causing the accelerated expansion of the universe. 

The gravitational lensing is a useful system to comprehend the galaxies, dark matter of the universe, dark energy and the universe \cite{6 18}. As the main gravitational lensing perception by the Eddington, a huge work on gravitational lensing have been accomplished for black holes, wormholes, cosmic strings and various types of spacetimes
(\cite{6 19}-\cite{6 33}). Since Einstein, geodesic technique (\cite{3 32}-\cite{3 37}) was considered to investigating the gravitational lensing and then in 2008, Gibbons and Werner showed new way to calculate weak deflection angle by introducing the Gauss-Bonnet theorem (GBT) \cite{3 38}. Using GBT, they find the light's weak gravitational deflection in the static and spherically symmetric (SSS) spacetime e.g. Schwarzschild spacetime \cite{3 38}. Then Werner showed that it is also possible to find weak deflection angle of stationary black holes such as Kerr black hole using GBT \cite{Werner:2012rc}. In GBT the light's deflection angle can be determined by integrating Gaussian curvature of related optical metric. In GBT we can utilize a space $D_{R}$, which is limited by the photon beam just as a circular boundary curve that is situated at focus on the focal point and photon beams meet the source and observer. It is expected that both origin and observer are at the coordinate distance R from the focal point. The GBT is written as \cite{3 38}:
\begin{equation}
    \int\int_{{D}_{R}} \mathcal{K}dS +\oint_{{\partial D}_{R}} \kappa~dt+\Sigma_{i} \theta_{i} =2\pi \mathcal{X}(D_{R}).\nonumber\\
\end{equation}
Here optical Gaussian curvature is denoted by $\mathcal{K}$ and an areal component is denoted by dS.
Subsequently thinking about the Euler characteristic. $\mathcal{X}(D_{R})=1$, also the jump angles are $\Sigma_{i} \theta_{i}=\pi$, by using straight light approximation and GBT, then the weak deflection angle is calculated by follows \cite{3 38}:
\begin{equation}
    \alpha=-\int_{0}^{\pi} \int_{\frac{b}{r\sin\phi}}^{\infty} \mathcal{K} dS.\nonumber\\
\end{equation}
Note that deflection angle is denoted by $\alpha$. Weak deflection angle using the GBT for various spacetimes were studied by many physicists. For example, the deflection angle of light studied for BHs and wormholes by the following authors (\cite{3 43}-\cite{Pantig:2020odu}), Ovgun et al. studied for different spacetimes like Schwarzschild-like spacetime in bumblebee gravity model (\cite{3 49}-\cite{3 51}) and  Javed et al. studied the impact of various matter fields (\cite{3 52}-\cite{3 55}). Next, Ishihara et al. \cite{1J 18} showed that it is conceivable to calculate weak deflection angle using the finite-distances method. Moreover, Ono et al. \cite{Ono:2017pie} has extended the method to stationary axisymmetric spacetimes. The strong deflection angle for finite distance has been discussed by Ishihara et al. \cite{Ishihara:2016sfv}. After that Crisnejo and Gallo 
\cite{1J 19} and many other authors (\cite{47}-\cite{65}) have contemplated the light's deflection angle in the presence of plasma medium.

Most of the scientist said that a supermassive BHs exist at the center of the Milky Way galaxy and they hope to detect its shadow and also the new results can give physicist an unprecedented look at black hole dynamics that will enable scientists to test general relativity (\cite{Akiyama:2019cqa}-\cite{S1k 47}). Some say that what we called a shadow is a dark interior while bright ring of radiation emitted by fast-moving, superheated gas swirling around, and falling into, the black hole. The shadow of a black hole is caused by gravitational light deflection.  Trajectory of a photon in vacuum is determined by its impact parameter. Photon sphere plays
crucial role in formation of the shadow. The size of the shadow radius for the black hole at the center of Milky way galaxy is about 53 $\mu as$. Shadow of the stationary black hole is deformed and oblate, while the shadow of the non-rotating black hole is just a circle. There are various types of scientific studies on black hole shadow in literature. For example, the shadow for negative tidal charges and charges corresponding to naked singularities are also discuss in \cite{Zakharov:2014lqa}. Moreover, Zakharov has studied constraints on a charge in the Reissner-Nordstr\"om metric for the black hole at the Galactic Center \cite{Zakharov:2014lqa}, and then has showed constraints on tidal charge of the supermassive black hole at the Galactic Center with trajectories of bright stars \cite{Zakharov:2018awx}.
Neves also has studied constraining the tidal charge of brane black holes using their shadows \cite{Neves:2020doc}.  Recently, Kocherlakota et al. have presented constraints on black hole charge from observations of $M87^{*}$ in 2017 \cite{EventHorizonTelescope:2021dqv}, where the authors used a dependence of shadow size on black hole charge for the Reissner–Nordstrom metric. It was shown that shadow size is decreasing with increasing a charge \cite{EventHorizonTelescope:2021dqv,Zakharov:2021gbg}. Based on results of shadow evaluation for $M87^{*}$ done by the EHT team, Zakharov constrains a tidal charge $q > -1.22$ as well as he evaluates a tidal charge $q> -0.25$ from shadow size estimates for $Sgr A^{*}$ in \cite{Zakharov:2021gbg}.

The paper is organized as follows: in Section II we derive the optical metric for the TCBH and find the weak deflection angle using the GBT in Section III. In Section IV we discus the graphical behavior and find the weak deflection angle in the presence of plasma medium using the GBT in Section V. Next, we find the graphical behavior in the presence of plasma in Section VI and calculate null geodesics of the TCBH in Section VII. Further we calculate shadow of the TCBH in Section VIII and also calculate shadow of the TCBH in the presence of plasma in Section IX and finally we conclude our results in Section X.

\section{Optical metric of TCBH}
The line element of a SSS TCBH is given by \cite{Dadhich:2000am,Pradhan:2014tva}
\begin{equation}
ds^2=-B(r)dt^2+ \frac{dr^2}{B(r)}+r^2d\Omega^2_{2},\label{AH0}
\end{equation}
where
\begin{equation}
B(r)=1-\frac{2M}{M_{p}^{2}r}+\frac{q}{M_{5}^{2}r^2},
\end{equation}
and
\begin{equation}
~~~d\Omega_{2}^{2}=d\theta^2+\sin^2\theta d\phi^2,
\end{equation}
here BH's mass is denoted by $M$ and the dimensionless tidal charge is denoted by $q$ and $M_{p}(=1.2 \times 10^{16} Tev)$ denotes the effective Planck mass on the brane and $M_{5}$ denotes the fundamental Planck scale in the 5D bulk and it is noted that generally $M_{5}<<M_{p}$.

By assuming equatorial coordinate plane $(\theta=\frac{\pi}{2})$, we obtain the optical metric of TCBH:
\begin{equation}
dt^2=\frac{dr^2}{B^{2}(r)}+\frac{r^2d\phi^2}{B(r)} .\label{H1}
\end{equation}

Then we calculate Gaussian optical curvature as follows:
\begin{equation}
\mathcal{K}=\frac{RicciScalar}{2} \approx -\frac{-2M}{r^{3}M_{p}^{2}}
-\frac{6qM}{r^{5}M_{p}^{2}M_{5}^{2}}+\frac{3q}{r^{4}M_{5}^{2}}
+\mathcal{O}\left(M^2,q^2\right).\label{AH1}
\end{equation}

\section{Deflection angle of TCBH}

In this section we find the deflection angle by TCBH with the help of GBT which is written as follows \cite{3 38}:
\begin{equation}
\int\int_{\mathcal{F}_{T}}\mathcal{K}dS+\oint_{\partial\mathcal{F}_{T}}kdt
+\sum_{l}\epsilon_{l}=2\pi\mathcal{Z}(\mathcal{F}_{T}),\label{AH6}
\end{equation}
here  $\mathcal{K}$ denote the Gaussian curvature and $k$ denote the geodesic curvature as;
$k=\bar{g}(\nabla_{\dot{\beta}}\dot{\beta},\ddot{\beta})$ so, $\bar{g}
(\dot{\beta},\dot{\beta})=1$, $\ddot{\beta}$ shows unit acceleration vector
and the $\epsilon_{l}$ shows the exterior angle at the lth vertex.
Since $T\rightarrow\infty$, both of the jump angles become $\pi/2$ and then we have $\theta_{O}+\theta_{T}\rightarrow\pi$. Euler characteristic is
$\mathcal{Z}(\mathcal{F}_{T})=1$, as $\mathcal{F}_{T}$ is non singular. So,
\begin{equation}
\int\int_{\mathcal{F}_{T}}\mathcal{K}dS+\oint_{\partial
\mathcal{F}_{T}}kdt+\epsilon_{l}=2\pi\mathcal{Z}(\mathcal{F}_{T}),
\end{equation}
here, $\epsilon_{l}=\pi$ demonstrates that both
$\alpha_{\bar{g}}$ and the total jump angles are geodesic and $\mathcal{Z}$ is the Euler characteristic number and it is equal to $1$. As
$T\rightarrow\infty$, then we have $k(E_{T})=\mid\nabla_{\dot{E}_{T}}\dot{E}_{ST\mid}$. Since, radial component
of geodesic curvature is written as \cite{3 38}
\begin{equation}
(\nabla_{\dot{E}_{T}}\dot{E}_{T})^{r}=\dot{E}^{\phi}_{T}
\partial_{\phi}\dot{E}^{r}_{T}+\Gamma^{0}_{11}(\dot{E}^{\phi}_{T})^{2}.\label{AH5}
\end{equation}
For large $T$, $E_{T}:=r(\phi)=T=const$. Hence, the equation
Eq.(\ref{AH5}) becomes $(\dot{E}^{\phi}_{T})^{2}=\frac{A^2(r)B(r)}{r^2}$.
As $\Gamma^{0}_{11}=-rB+\frac{r^{2}B'(r)}{2}$, so it becomes
\begin{equation}
(\nabla_{\dot{E}^{r}_{T}}\dot{E}^{r}_{T})^{r}\rightarrow\frac{1}{T},
\end{equation}
so that $k(E_{T})\rightarrow T^{-1}$.
By using the optical metric Eq.(\ref{H1}) and write it as $dt=Td\phi$, we have $k(E_{T})dt=d\phi$,

By combining all the result we have,
\begin{equation}
\int\int_{\mathcal{F}_{T}}\mathcal{K}ds+\oint_{\partial \mathcal{F}_{T}} kdt
=^{T\rightarrow\infty}\int\int_{U_{\infty}}\mathcal{K}dS+\int^{\pi+\Theta}_{0}d\phi.\label{hamza2}
\end{equation}

Light ray in weak field limits at zeroth order (straight light approximation) is written as $r(t)=b/\sin\phi$.
Using the above steps, the weak deflection angle can be calculated by \cite{3 38}
\begin{equation}
\Theta=-\int^{\pi}_{0}\int^{\infty}_{b/\sin\phi}\mathcal{K}\sqrt{det\bar{g}} ~dr d\phi,\label{AH7}
\end{equation}
here $\sqrt{det\bar{g}}\approx rdr$.

After replacing the Gaussian optical curvature
Eq.(\ref{AH1}) into Eq.(\ref{AH7}), the weak deflection angle is obtained as:
\begin{eqnarray}
\Theta &\thickapprox& \frac{4M}{b M_{p}^2}-\frac{3q\pi}{4b^{2}M_{5}^{2}}+\mathcal{O}(M^2,q^2).\label{P1}
\end{eqnarray}

On the other hand, deflection angle of Reissner-Nordstrom (RN) BH in \cite{RN} is given as;
\begin{equation} \label{P1}
\Theta_{RNBH} \thickapprox \frac{4M}{b}+\frac{15\pi M^2}{4b^2}-\frac{3q^2\pi M^2}{4b^2}+\mathcal{O}(M^3,b^3).
\end{equation}
Note that increasing the value of tidal charge decreases the deflection angle. We compare these results in the graphical section.

\section{Graphical Analysis for non-plasma medium}
In this section we obtain the graphical behavior of TCBH's deflection angle. We check the graphical behavior in correspondence of deflection angle $\Theta$ with impact parameter $b$ by varying the value of dimensionless tidal charge $q$. 

\begin{figure}[ht!]
   \centering
    \includegraphics[scale=0.7]{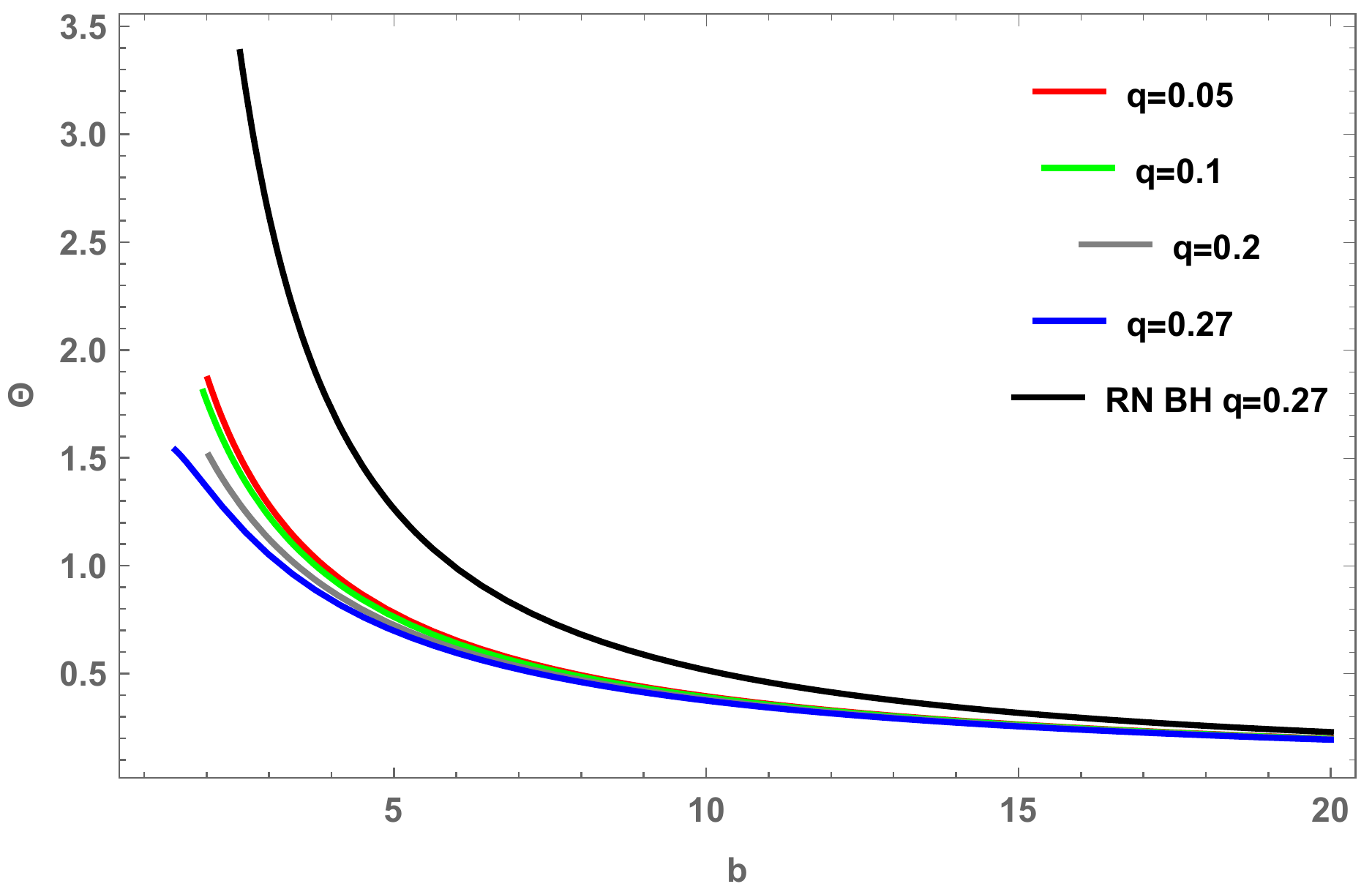}
    \caption{Relation between $\Theta$ and $b$ of TCBH for positive values of $q$ for $M_p=1$, $M_{5}=0.5$, $M=1$.}
    \label{fig:lensing1}
\end{figure}

\begin{figure}[ht!]
   \centering
    \includegraphics[scale=0.9]{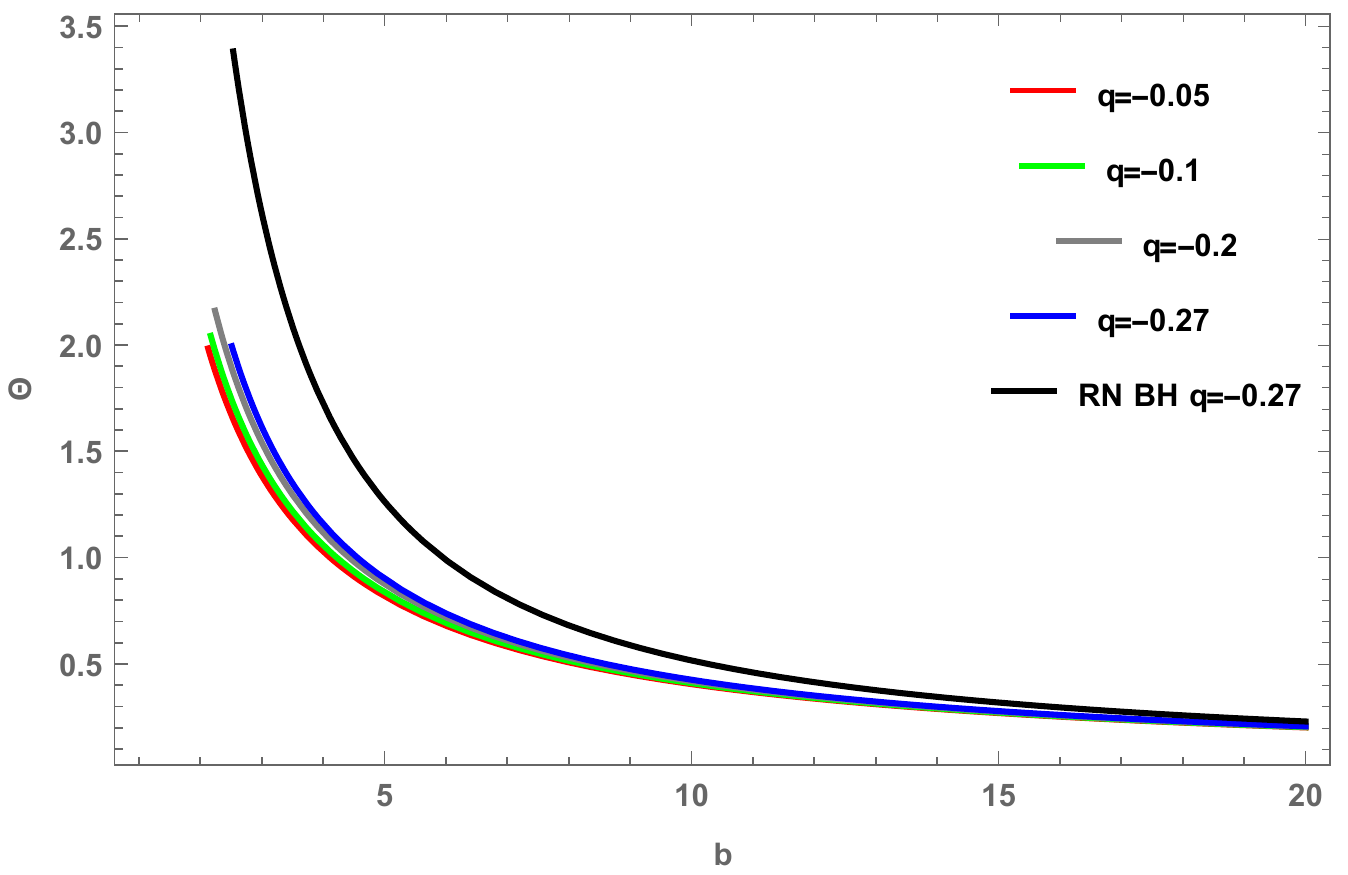}
    \caption{Relation between $\Theta$ and $b$ of TCBH for negative values of $q$ for $M_p=1$, $M_{5}=0.5$, $M=1$.}
    \label{fig:lensing2}
\end{figure}

\textbf{Figure \ref{fig:lensing1} and Figure \ref{fig:lensing2}} demonstrate the relation of $\Theta$ w.r.t $b$ for different values of $q$ of TCBH. We observe that deflection angle continuously decreases for increasing of $q$ and shows a constant behavior.

By comparing the graphs of RNBH and TCBH, we observe that deflection angle of RN has larger than the deflection angle of TCBH. We conclude that the tidal charge decreases the deflection angle as compared with RN charge in Fig. \ref{fig:lensing1} and \ref{fig:lensing2}.

\section{Effect of plasma on gravitational lensing}
In this section, we check how a plasma medium affects the gravitational lensing of TCBH. Now, we assume the TCBH in the presence of plasma illustrated by the refractive index $n$, (\cite{51}-\cite{53})
\begin{equation}
 n^2\left(r,\omega(r)\right)=1-\frac{\omega_e^2(r)}{\omega^2(r)}.
\end{equation}
Here, refractive index is defined as follows;
\begin{equation}
n(r)=\sqrt{{1-\frac{\omega_e^2}{\omega_\infty^2}\left(1-\frac{2M}{M_{p}^{2}r}+\frac{q}{M_{5}^{2}r^2}\right)}},
\end{equation}
where  $\omega(r)$ is the photon frequency and $\omega_e(r)$ is the electron plasma frequency
and $\omega_{e}$ is considered as homogeneous plasma. 
The optical metric of Eq. \ref{AH0} in plasma medium, is written as \cite{1J 19}
\begin{equation}
 dt^2=g^{opt}_{lm}dx^ldx^m=n^2 \left[\frac{dr^2}{B^2(r)}+\frac{r^2d\phi^2}{B(r)}\right],\label{hamza3}
\end{equation}
 with determinant $g^{opt}_{lm}$,
\begin{equation}
 \sqrt{g^{opt}}=r(1-\frac{\omega_e^2}{\omega_\infty^2})+\frac{M}{M_{p}^{2}(r)}(3
 +\frac{\omega_e^2}{\omega_\infty^2})-\frac{q}{2M_{5}^{2}r}(3
 +\frac{\omega_e^2}{\omega_\infty^2}).
\end{equation}
 
Then the optical Gaussian curvature can be calculated by follows:
\begin{eqnarray}
\mathcal{K}&=&-3\,{\frac {M{\omega_e}^{2}}{{r}^{3}{\omega_\infty}^{2}{{
\it M_{p}}}^{2}}}-2\,{\frac {M}{{{ M_{p}}}^{2}{r}^{3}}}+5\,{\frac {q{
\omega_e}^{2}}{{\omega_\infty}^{2}M_{5}^{2}{r}^{4}}}+3\,{
\frac {q}{M_{5}^{2}{r}^{4}}}-26\,{\frac {qM{\omega_e}^{2}}{{
\omega_\infty}^{2}M_{5}^{2}{{ M_{p}}}^{2}{r}^{5}}}-6\,{\frac {qM}{M_{5}
^{2}{{M_{p}}}^{2}{r}^{5}}}.
\end{eqnarray}
With the help of GBT we obtained the deflection angle of TCBH in the presence of plasma medium. So, for obtaining deflection angle in weak field limit, we apply the condition of $ r=\frac{b}{sin\phi}$ at 0th order:
\begin{equation}
    \Theta=-\lim_{R\rightarrow 0}\int_{0} ^{\pi} \int_\frac{b}{\sin\phi} ^{R} \mathcal{K} dS,
\end{equation}
  using the above equation with the optical Gaussian curvature, the deflection angle of light in the presence of plasma medium is found as follows:
\begin{eqnarray}
\Theta&=&
4\,{\frac {M}{b{{\it M_{p}}}^{2}}}+3/4\,{\frac {q{\omega_e
}^{2}\pi}{{b}^{2}{\omega_\infty}^{2}{M_{5}}^{2}}}-3/4\,{\frac {q\pi}{{b
}^{2}{M_{5}}^{2}}}-6\,{\frac {M{\omega_e}^{2}}{b{\omega_\infty}^{2}{{\it M_{p}}}
^{2}}}+\mathcal{O}(M^2,q^2,\frac{\omega_e^3}{\omega_\infty^3}).\label{P2}
\end{eqnarray}

Note that one can neglect the plasma medium effect by $(\frac{\omega_e}{\omega_\infty}\rightarrow0)$ then this deflection angle in Eq. $(\ref{P2})$ reduces into the angle in Eq. $(\ref{P1})$.

\section{Graphical Analysis for plasma medium}
 Here, for simplicity we take $\frac{\omega_e}{\omega_\infty}$=$10^{-1}$ and observe the deflection angle by changing the value of dimensionless tidal charge $q$. 

\begin{figure}[ht!]
   \centering
    \includegraphics[scale=0.7]{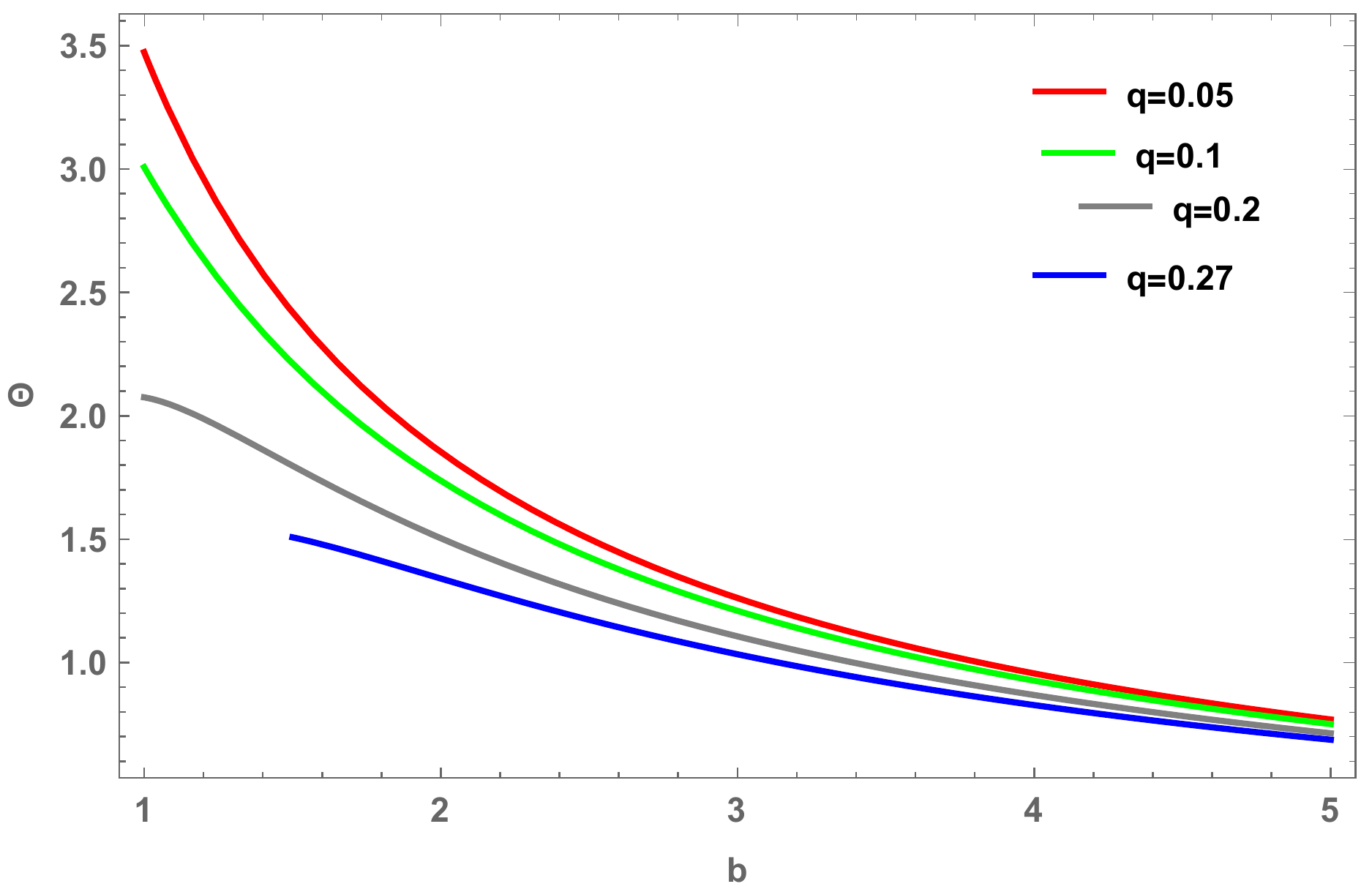}
    \caption{Relation between $\Theta$ and $b$ of Tidal Charged BH for positive values of $q$ for $M_p=1$, $M_{5}=0.5$, $M=1$ and $\frac{\omega_e}{\omega_{\infty}}=10^{-1}$ for exaggeration of the plot.}
    \label{fig:lensing3}
\end{figure}

\textbf{Figure \ref{fig:lensing3}} demonstrates the relation of $\Theta$ with $b$ for different values of $q$ in plasma medium. In this plot, we observe that for increasing the value of $q$, decreases the deflection angle in plasma medium. Moreover, the deflection angle in plasma medium is smaller that the deflection angle in vacuum.

\section{Null Geodesic in a TCBH}
The Lagrangian representing the motion of light in the
TCBH's spacetime using the Eq.(\ref{AH0}) is written as;
\begin{equation}
2\mathcal{L}=-\left(1-\frac{2M}{M_{p}^{2}r}+\frac{q}{M_{5}^{2}r^2}\right) \dot t^2+
\left(1-\frac{2M}{M_{p}^{2}r}+\frac{q}{M_{5}^{2}r^2}\right)\dot r^2 +r^2\dot \theta^2+r^2\sin^2\theta \dot \phi^2,
\end{equation}
here derivative w.r.t affine parameter $\lambda$ is represented by an overdot. As Lagrangian is not depend on t and $\phi$ so we introduce two new constants named as energy $E$ and the angular momentum $L$ where the values of these constants is defined as;
\begin{equation}
p_{t}=\frac{\partial L}{\partial \dot t}=-(1-\frac{2M}{M_{p}^{2}r}+\frac{q}{M_{5}^{2}r^2})\dot t=-E,
\end{equation}
and
\begin{equation}
p_{\phi}=\frac{\partial L}{\partial \dot \phi}=r^2\sin^2\theta \dot \phi=L.
\end{equation}

For finding the restrictions of geodesic, we use these constants as;
\begin{equation}
\frac{dt}{d\lambda}=\dot t=\frac{E}{1-\frac{2M}{M_{p}^{2}r}+\frac{q}{M_{5}^{2}r^2}}~~~,~~~
\frac{d\phi}{d\lambda}=\dot \phi=\frac{L}{r^2\sin^2\theta}.
\end{equation}
Now, we define the new parts of momentum named as $r$-part and $\theta$-par;
\begin{equation}
p_{r}=\frac{\partial L}{\partial \dot r} =\frac{\dot r}{1-\frac{2M}{M_{p}^{2}r}+\frac{q}{M_{5}^{2}r^2}}~~and~~
p_{\theta}=\frac{\partial L}{\partial \dot \theta}=r^2\dot \theta.
\end{equation}
With the help of Hamilton-Jacobi equation, we find the values of $r$-part and $\theta$-part of the geodesic equation as;
\begin{equation}
\frac{\partial S}{\partial \lambda}=-\frac{1}{2}g^{\mu \nu}\frac{\partial S}
{\partial x^\mu}\frac{\partial S}{\partial x^\nu},\label{S1}
\end{equation}
and for photons $(m_0 = 0)$,  Eq. (\ref{S1}) can give us result of the following type,
\begin{equation}
S=-Et+L\phi+S_r(r)+S_\theta(\theta),\label{S2}
\end{equation}
where $S_r$ depends on $r$ and $S_\theta$ depends on $\theta$. Now we get the Carter constant $(\pm\mathcal{K})$ \cite{my2 SA92} by separating the values of $r$ and $\theta$ and we get these values of $r$ and $\theta$ by replacing  Eq. (\ref{S2}) into Eq. (\ref{S1}) and also substituting the values of contravariant metric, i.e., $g^{\mu\nu}$, we have
\begin{eqnarray}
\frac{1}{\sqrt{1-\frac{2M}{M_{p}^{2}r}+\frac{q}{M_{5}^{2}r^2}}}\frac{dr}{d\lambda}=^+_-\sqrt R(r),\nonumber\\
r^2\frac{d\theta}{d\lambda}=^+_-\sqrt T(\theta),
\end{eqnarray}
here the values of $R$ and $\theta$ can be defined as,
\begin{eqnarray}
R(r)=\frac{E^2}{1-\frac{2M}{M_{p}^{2}r}+\frac{q}{M_{5}^{2}r^2}}-\frac{\mathcal{K}}{r^2},\nonumber\\
T(\theta)=\mathcal{K}-\frac{L^2}{\sin^2\theta}.
\end{eqnarray}
Now, Equation $S_r$ can be written as
\begin{equation}
\frac{dr}{d\lambda}^2+V_{eff}=0,
\end{equation}
with
\begin{equation}
V_{eff}=-\left({1-\frac{2M}{M_{p}^{2}r}+\frac{q}{M_{5}^{2}r^2}}\right)R(r).
\end{equation}
We see that effective potential depend on  BH's mass denoted by $M$ and the dimensionless tidal charge denoted by $q$ and effective Planck mass on the brane denoted by $M_{p}$  and  fundamental Planck scale in the 5D bulk denoted by $M_{5}$ and radius $r$ and $R(r)$. Now we change these parameters to new impact parameters such as $\xi =\frac{L}{E}$ and $\eta=\frac{\mathcal{K}}{E^2}$. Now we change the value of $R$ with respect to these new impact parameters.
\begin{equation}
R=E^2[\frac{1}{1-\frac{2M}{M_{p}^{2}r}+\frac{q}{M_{5}^{2}r^2}}-\frac{\eta}{r^2}].\label{S3}
\end{equation}
\section{Shadow of TCBH}
Here, in this section we find the shadow of TCBH and we discussed in detail about shadow in the introduction. Now for finding the shadow, we find the unstable circular photons orbits \cite{Synge:1966okc}. For this we must satisfy this,
\begin{equation}
R=0~~~~~and~~~~~R^{\prime}=0,\label{S4}
\end{equation}
where prime $(\prime)$ means differentiation w.r.t $r$. Putting (\ref{S3}) into (\ref{S4}), we obtained the relation for photon sphere:
\begin{equation}
\frac{B^{\prime}(r)}{B(r)}=\frac{2}{r}.
\end{equation}

and the photon sphere $r_p$ derived as follows:

\begin{equation}
r_p=\frac{3 M_{5}^2 M+\sqrt{9 M_{5}^4 M^2-8 M_{5}^2 M_{p}^4 q}}{2 M_{5}^2 M_{p}^2}. \end{equation}

The radius of the shadow $r_s$ at photon sphere radius $r_p$ is calculated as follows:

\begin{equation}
r_s=\sqrt{\eta+\xi^{2}}=\frac{r_p}{\sqrt{1-\frac{2M}{M_{p}^{2}r_p}+\frac{q}{M_{5}^{2}r_p^2}}},\label{S5}
\end{equation}

\begin{equation}
r_s=\frac{1}{2} \sqrt{\frac{\left(3 M_{5}^2 M+\sqrt{9 M_{5}M_{5}^4 M^2-8 M_{5}^2 M_{p}^4 q}\right)^3}{M_{5}^4 M_{p}^2 \left(3 M_{5}^2 M^2+M \sqrt{9 M_{5}^4 M^2-8 M_{5}^2 M_{p}^4 q}-2 M_{p}^4 q\right)}},
\end{equation}
where impact parameters depends on BH's mass denoted by $M$ and the dimensionless tidal charge denoted by $q$ and effective Planck mass on the brane denoted by $M_{p}$ and fundamental Planck scale in the 5D bulk denoted by $M_{5}$. So Eq. (\ref{S5}) gives detail about the boundary of the shadow and an observer which is far away from the BH can find this shadow in this sky and we make new coordinates in the observer's sky named as the celestial coordinates ($\alpha$,$\beta$), and we relate these coordinates with impact parameters ($\xi$,$\eta$). These coordinates are defines in these papers (\cite{SA47},\cite{SA48}) as;
\begin{eqnarray}
\alpha=^{\lim}_{r_{0} \rightarrow \infty}(r_0^2\sin\theta_0)\frac{d\phi}{dr},\nonumber\\
\beta=^{\lim}_{r_{0} \rightarrow \infty} ~ r_{0}^{2}\frac{d\theta}{dr}.\label{S6}
\end{eqnarray}
Note that $r_0$ represents the distance between the viewer and the BH and $\theta_0$ denotes the angular coordinates of the observer called "inclination angle". After putting the equations of four-velocities into Eq. (\ref{S6}), and doing some calculation, we obtain these celestial coordinates as,
\begin{equation}
\alpha=-\frac{\xi}{\sin\theta_0}~~~~~~~and~~~~~~~ \beta=\sqrt{\eta-\frac{\xi^2}{\sin\theta_{0}^{2}}}.
\end{equation}
With the help of these equations and using the impact parameters we now make the shape of TCBH's shadow and for plotting the shadows's shape we plot  $\alpha$ versus $\beta$ which give detail about the boundary of the TCBH in the observer's sky and these plots are seen in the Table I and Fig 4.

We plots the shadow of the black hole by changing the values of the dimensionless tidal charge $q$ using the constraints from the astronomical observations on the upper limiting values on
tidal charge parameters \cite{Zakharov:2018awx,Neves:2020doc}, in this plot we also discuss about the shadow for positive and negative values of $q$. We see that shadow's shape is a perfect circle and its shape is shown in Fig. 4. In this graph we see that for small values of $q$, radius of the shadow shows different behavior. The increasing the value of the tidal charge decrease the radius of the shadow as well as radius of the photon sphere.

	\begin{table}[ht!]
    \centering
    \begin{tabular}{ |p{1cm}||p{2cm}|p{2cm}| }
    \hline
        $q$ &  $r_{p}$ &  $R_{s}$ \\ [0.5ex] 
        \hline
        0.05 & 2.86015& 2.96572  \\
        0.1 & 2.70416& 2.92949 \\
        0.2 &2.30623& 2.8537 \\
        0.27 & 1.8   & 2.84605 \\
        \hline
    \end{tabular}
    \caption{Effects of the tidal charge on the BH shadow for fixed $M_p=1$, $M_{5}=0.5$, $M=1$.}
    \label{table1}
\end{table}

\begin{figure}[ht!]
   \centering
    \includegraphics[scale=0.5]{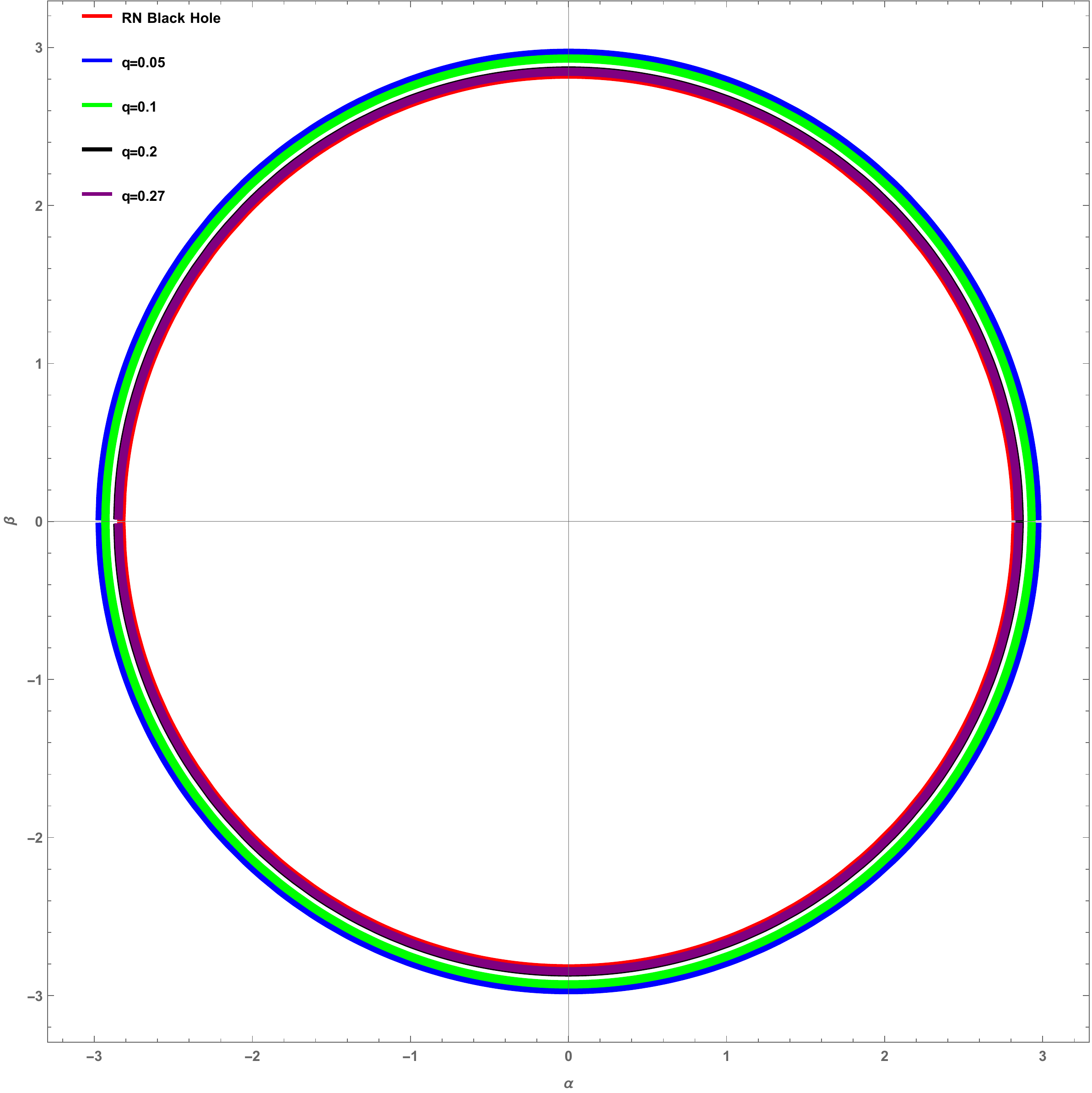}
    \caption{Shadow of the black hole for different values of $q$ for $M_p=1$, $M_{5}=0.5$, $M=1$.}
    \label{fig:shadow}
\end{figure}

Here, we consider the reported angular size of the shadow of the $M87^{*}$ black hole by EHT  as $\theta_s = (42 \pm 3)\mu as$, where the distance to $M87^{*}$ is $D = 16.8 $ Mpc, as well as the mass of $M87^{*}$ is  $M = 6.5 \times 10^9$ M\textsubscript{\(\odot\)} \cite{Akiyama:2019cqa}. Then, the diameter of the shadow in units of mass $d_{M87^{*}}$ is given by \cite{Allahyari:2019jqz}
\begin{eqnarray}
d_{M87^{*}}=\frac{D \,\theta_s}{M87^{*}}=11.0 \pm 1.5.
\end{eqnarray}

As compared the radius of the shadow   $d_{M87}$ black hole  $r_{M87^{*}}=\frac{d_{M87^{*}}}{2}$ with the radius of the shadow of tidal charged black hole $R_s$ in Table I. It is clear that the shadow of the tidal charge black hole has smaller radius of the shadow.

\section{Effect of Plasma on Shadow of TCBH}
In Paper \cite{S1k 47}, authors discuss in detail about calculating the shadow of a spherically symmetric spacetime in plasma medium. The line element of spherically symmetric TCBH is defined in Eq.(\ref{AH0}) as follows
\begin{equation}
ds^2=-A(r)dt^2+ \frac{dr^2}{B(r)}+r^2d\Omega^2_{2},
\end{equation}
 where
\begin{equation}
A(r)=[B(r)]^{-1}=1-\frac{2M}{M_{p}^{2}r}+\frac{q}{M_{5}^{2}r^2},
\end{equation}
and here we check the TCBH's shadow in the presence of plasma medium. The refractive index $n$ can be written as;
\begin{equation}
 n^2\left(r,\omega(r)\right)=1-\frac{\omega_e^2(r)}{\omega^2(r)}.
\end{equation}
Here from reference\cite{S1k 47}, we can define the $h(r)$ as;
\begin{equation}
h(r)^2=r^2\left(\frac{1}{1-\frac{2M}{M_{p}^{2}r}+\frac{q}{M_{5}^{2}r^2}}-\frac{\omega_e^2(r)}{\omega_\infty^2}\right).\label{ss1}
\end{equation}
We can calculate the radius of a photon sphere by computing the above equation as,
\begin{equation}
0=\frac{d}{dr}\left(h{(r)}^2\right)\label{ss2}.
\end{equation}
In this case the value of $h(r)$ is defined by (\ref{ss1}) and we use circumstances (\ref{ss2}) for a photon sphere becomes
\begin{equation}
0=\frac{4r^3(r^2-\frac{2Mr}{M_{p}^{2}}+\frac{q}{M_{5}^{2}})-r^4}{\left(r^2-\frac{2Mr}{M_{p}^{2}}+\frac{q}{M_{5}^{2}}\right)^2}
-2r\frac{\omega_e^2(r)}{\omega_\infty^2}-2r^2\frac{\omega_e(r)\omega_e(r')}{\omega_\infty^2}.\label{ss3}
\end{equation}
Angular radius of the shadow is defined as,
\begin{equation}
\sin^{2}\alpha_{sh}=\frac{h(r_{ph})^2}{h(r_O)^2}.\label{ss4}
\end{equation}
Now, after putting the value of Eq. (\ref{ss1}) into the Eq. (\ref{ss4}) we have

\begin{equation}
\sin^{2}\alpha_{sh}=\frac{r_{ph}^2\left(\frac{1}{1-\frac{2M}{M_{p}^{2}r_{ph}}+\frac{q}{M_{5}^{2}r_{ph}^{2}}}-\frac{\omega_e^2(r_{ph})}{\omega_\infty^2}\right)}
{r_{O}^{2}\left(\frac{1}{1-\frac{2M}{M_{p}^{2}r_{O}}+\frac{q}{M_{5}^{2}r_{O}^{2}}}-\frac{\omega_e^2(r_{O})}{\omega_\infty^2}\right)},
\end{equation}
where $r_{ph}$ has to be determined by putting Eq. (\ref{ss1}) into Eq. (\ref{ss2}). For vacuum, $\omega_e(r)=0$, our consideration
gives
\begin{equation}
h(r)^2=r^2\left(\frac{1}{1-\frac{2M}{M_{p}^{2}r}+\frac{q}{M_{5}^{2}r^2}}\right)
\end{equation}
and
\begin{equation}
\sin^{2}\alpha_{sh}=\frac{r_{ph}^2\left(\frac{1}{1-\frac{2M}{M_{p}^{2}r_{ph}}+\frac{q}{M_{5}^{2}r_{ph}^{2}}}\right)}
{r_{O}^{2}\left(\frac{1}{1-\frac{2M}{M_{p}^{2}r_{O}}+\frac{q}{M_{5}^{2}r_{O}^{2}}}\right)},
\end{equation}
where the positive value of $r_{ph}$ can be calculated as;
\begin{equation}
r_{ph}=\frac{1+\frac{2M}{M_{p}^{2}} {+}\sqrt{(1+\frac{2M}{M_{p}^{2}})^2-\frac{4q}{M_{5}^{2}}}}{2}.
\end{equation}

\section{Conclusion}
In this paper, we first calculate the weak deflection angle of TCBH with the help of Gaussian curvature. To do so, We use the GBT for calculating the weak deflection angle which was first proposed by Gibbons and Werner. The deflection angle is found as follows:
\begin{eqnarray}
\Theta &\thickapprox& \frac{4M}{b M_{p}^2}-\frac{3q\pi}{4b^{2}M_{5}^{2}}+\mathcal{O}(M^2,q^2).\label{P11}
\end{eqnarray}

This shows that weak deflection angle depends on BH's mass denoted by $M$, the dimensionless tidal charge denoted by $q$, effective Planck mass on the brane denoted by $M_{p}$, fundamental Planck scale in the 5D bulk denoted by $M_{5}$ and impact parameter $b$. After calculating the deflection angle of TCBH we check its graphical behavior by varying the values of $q$ and by fixing all the other constants. Moreover, we study the weak deflection angle of TCBH in the presence of plasma medium using the GBT. This angle in the presence of plasma is obtained as follows:
\begin{eqnarray}
\Theta&=&
-6\,{\frac {M{\omega_e}^{2}}{b{\omega_\infty}^{2}{{\it M_{p}}}
^{2}}}-4\,{\frac {M}{b{{\it M_{p}}}^{2}}}+5/4\,{\frac {q{\omega_e
}^{2}\pi}{{b}^{2}{\omega_\infty}^{2}{M_{5}}^{2}}}+3/4\,{\frac {q\pi}{{b
}^{2}{M_{5}}^{2}}}+\mathcal{O}(M^2,q^2,\frac{\omega_e^3}{\omega_\infty^3}). \label{P22}
\end{eqnarray}

After neglecting the plasma medium effect we find the same angle as we find in the non-plasma case. Now if we neglect the plasma effect $(\frac{\omega_e}{\omega_\infty}\rightarrow0)$ then this deflection angle Eq. $(\ref{P22})$ reduces into this angle Eq. $(\ref{P11})$. This shows the correctness of our angle in the presence of plasma medium. After calculating the effect of plasma, we find the graphical behavior of TCBH in the presence of plasma medium. It does not shows the same behavior as the behavior without plasma. Hence, we show that deflection angle continuously decreases for increasing of $q$  in Figure \ref{fig:lensing1} and Figure \ref{fig:lensing2}. Moreover, in Figure \ref{fig:lensing3} we demonstrate that for increasing the value of $q$, decreases the deflection angle in plasma medium as well as the deflection angle in plasma medium is smaller that the deflection angle in vacuum.

Last, we also find the shadow of TCBH by studying the null geodesic of TCBH. After that we calculate the radius of the shadow and show its image in the far away observer's sky, using the celestial coordinates ($\alpha$,$\beta$). Hence, we show that the shadow of the black hole by changing the values of the dimensionless tidal charge $q$ using the constraints from the astronomical observations on the upper limiting values on
tidal charge parameters \cite{Zakharov:2018awx,Neves:2020doc}, in the Fig. \ref{fig:shadow}, we also discuss about the radius of the shadow for positive and negative values of $q$. We see that shadow's shape is a perfect circle. We conclude that  for small values of $q$, radius of the shadow shows different behavior and the increasing the value of the tidal charge decrease the radius of the shadow as well as radius of the photon sphere. Hence, we show that the shadow of the tidal charge black hole has smaller radius of the shadow as compared with the $M87^{*}$ black hole recently observed by EHT.

\end{document}